\documentclass[9pt,twocolumn,twoside]{pnas-new}

\usepackage{xspace}
\usepackage[normalem]{ulem}

\newcommand{\z}{3z$^2$-r$^2$}
\newcommand{\xy}{x$^2$-y$^2$}
\newcommand{\eg}{e$_g$}

\templatetype{pnasresearcharticle} 

\title{Disentangling lattice and electronic contributions to the metal–insulator transition from bulk vs. layer confined RNiO$_3$}

\author[1]{Alexandru B. Georgescu}
\author[2]{Oleg E. Peil} 
\author[3]{Ankit S. Disa}
\author[1,4,5,6]{Antoine Georges}
\author[1,7]{Andrew J. Millis}

\affil[1]{{Center for Computational Quantum Physics, Flatiron Institute, 162 5th Avenue, New York, NY 10010, USA}}
\affil[2]{Materials Center Leoben, Leoben, Austria}
\affil[3]{Max Planck Institute for the Structure and Dynamics of Matter, Hamburg, Germany}

\affil[4]{Coll\`ege de France, 11 place Marcelin Berthelot, 75005 Paris, France}
\affil[5]{Centre de Physique Th\'{e}orique Ecole Polytechnique, CNRS, Universite Paris-Saclay, 91128 Palaiseau, France}
\affil[6]{Department of Quantum Matter Physics, University of Geneva, 24 Quai Ernest-Ansermet, 1211 Geneva 4, Switzerland}
\affil[7]{Columbia University, NYC, NY, United States}

\leadauthor{Georgescu} 

\significancestatement{Our combined theoretical and experimental study of bulk and heterostructured forms of a correlated electron material leads to new insights into the metal-insulator transition. Comparison of single layer, bilayer and very thick samples validates a combined ab-initio/many-body theoretical approach and enables a clear disentangling of electronic and lattice contributions to the transition by independently changing each. Analysis of the lattice relaxations associated with the metal-insulator transition highlights the importance of the elastic properties of and propagation of distortions into the electronically inert counterlayer, defining a new control parameter for tuning electronic properties. Counterlayer induced bond angle changes and electronic confinement provide separate tuning parameters, with bond angle changes found to be a much less effective tuning parameter. 
}

\correspondingauthor{\textsuperscript{2}To whom correspondence should be addressed. E-mail: ageorgescu\@flatirioninstitute.org}


\begin{abstract} 
In complex oxide materials, changes in electronic properties are often associated with changes in crystal structure, raising the question of the relative roles of the electronic and lattice effects in driving the metal-insulator transition. This paper presents a combined theoretical and experimental analysis of the dependence of the metal-insulator transition of NdNiO$_3$ on crystal structure, specifically comparing properties of bulk materials to one and two layer samples of NdNiO$_3$ grown between multiple electronically inert NdAlO$_3$ counterlayers in a superlattice. The comparison amplifies and validates a theoretical approach developed in previous papers and disentangles the electronic and lattice contributions, through an independent variation of each. In bulk NdNiO$_3$ the correlations are not strong enough to drive a metal-insulator transition by themselves: a lattice distortion is required. Ultra-thin films exhibit two additional electronic effects and one lattice-related effect. The electronic effects are quantum confinement, leading to dimensional reduction of the electronic Hamiltonian, and an increase in electronic bandwidth due to counterlayer induced bond angle changes. We find that the confinement effect is much more important. The lattice effect is an increase in stiffness due to the cost of propagation of the lattice disproportionation into the confining material.

\end{abstract}

\dates{This manuscript was compiled on \today}
\doi{\url{www.pnas.org/cgi/doi/10.1073/pnas.XXXXXXXXXX}}

\begin{document}

\maketitle
\thispagestyle{firststyle}
\ifthenelse{\boolean{shortarticle}}{\ifthenelse{\boolean{singlecolumn}}{\abscontentformatted}{\abscontent}}{}


\subsection*{Introduction:} Metal insulator transitions (MIT) in correlated electron materials typically involve changes in both the electronic and atomic structure. The relative importance of the two effects has been the subject of extensive discussion\citep{Park2013,Park2014, Han2018,Haule2016,Haule2017Nickelates,Mandal2017,Amadon2012,Oleg}. In this paper, using a recently developed theoretical approach \cite{Han2018,Oleg}, we  argue that comparison of few-layer and bulk materials yields considerable insight into the relative importance of electronic and lattice contributions, essentially because these  are affected by heterostructuring in opposite ways. We disentangle these effects by independently changing each. Motivated by recent experimental \citep{Disa2017,Triscone2018,SohrabPicoscale,Superlattices2011,Kumah2014a,Middey2018,Gray2011,Stemmer2013,AlexEELS,Stemmer2013,Zhang2016b,Shamblin2018,Meyers2016,Forst2017,Forst2015,Medarde1998,Hu2016,Caviglia2013} and theoretical \citep{guzman,AndyArxiv,Park2016,OpticalAntoine,Kumah2014a,polarizationantoine,Han2011,Oleg, Seth2017, Subedi2015, OpticalAntoine, Strand2014,Blanca-Romero2011,held} results, we focus here on the rare earth nickelate family of materials. The concepts, formalism and findings are applicable to wide classes of materials.

\begin{figure}
\centering
\includegraphics[width=0.9\linewidth]{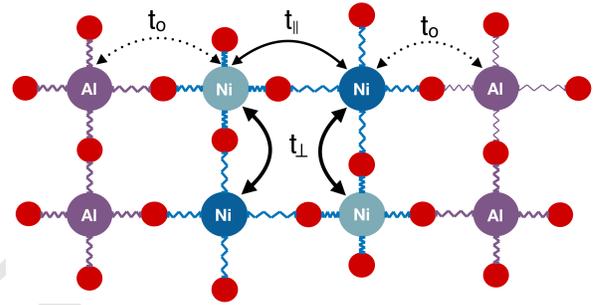}
\caption{\textbf: Heterostructuring NdNiO$_3$ with NdAlO$_3$: structural distortions are represented as motion (not to scale)  of oxygen ions (red circles) away from center of Ni-Ni and Ni-Al bonds; note that the distortions do not propagate significantly into the Al layers are are of reduced amplitudes along the Ni-Al bonds. The different colors of the Ni atoms represent the electronic disproportionation. The kinetic energy of the Ni $e_g$ electrons is reduced by confinement, as electrons are not allowed to hop through the insulating NAO layers (|t$_0$|<<|t$_\bot|$,|t$_\parallel$|), while |t$_\parallel$| < |t$_\perp$| due to propagation of bond angles from NdAlO$_3$.  }
\label{fig:frog}
\end{figure}

The rare earth nickelates have chemical formula RNiO$_3$ (R is a rare earth metal of the lanthanide rare earth series). In bulk, at high T, they are metallic and form an orthorhombic Pbnm structure (except for R=La for which the structure is rhombohedral) that is a distorted ABO$_3$ cubic perovskite in which the Ni ions are equivalent up to a rotation and translation. For all R except for La, the bulk materials undergo a metal to insulator transition (MIT) as T is decreased. The transition is first order and the low T phase has a P2$_1$/n structure with two fundamentally inequivalent Ni sites characterized by an electronic charge disproportionation $\Delta N$ and a lattice distortion Q, both defined more precisely below. The relative roles of the two has been the subject of debate. The issue has typically been addressed by calculations (typically performed at fixed crystal structure) and experiments on a specific material or on members of a family of materials, and has not been resolved. Similar issues arise in many other transition metal oxide materials.

Recent experiments \citep{Disa2017} report that  in NdNiO$_3$/NdAlO$_3$ (NNO/NAO) superlattices in which one or two monolayers of NNO are separated by many layers of the wide-gap insulator NAO, the MIT occurs at a much higher temperature than in the bulk, while the X-ray signatures of the lattice distortion are much  less pronounced in the superlattices than in bulk. These experiments suggest that heterostructuring affects electronic and lattice properties differently and thus that a comparative examination of the two material forms can help disentangle the relative importance of electronic and lattice contributions to the metal-insulator transition.  In this paper we theoretically investigate the differences between bulk NdNiO$_3$ and superlattice NdNiO$_3$/NdAlO$_3$ materials using a theoretical approach previously applied to bulk nickelates \citep{Oleg,Han2011,Park2012,Subedi2015,Seth2017,OpticalAntoine,AndyArxiv,Haule2017Nickelates,Mercy2017,Johnston2014b} and to ruthenates \citep{Han2018}. 

In figure \ref{fig:frog} we represent the main phenomenology that we disentangle in this paper, as exemplified on the bilayer NNO. Namely, the structural distortions in the material become inhomogenous due to the presence of the NdAlO$_3$ counterlayer and the absence of a driving force on the interfacial oxygen from the aluminum atom. In our effective model, this leads to an increased effective stiffnes of the bond disproportionation mode as the same force from the nickel atoms leads to a lower average oxygen displacement. On the electronic structure, the layer confinement of the material leads to suppressed hopping along the z direction, while the propagation of bond angles from NAO to NNO leads to a small increase of the Ni-Ni hopping in-plane as compared to the out-of-plane Ni-Ni hopping in the bilayer.

\subsection*{Energy} Central to our discussion is an expression for the energy difference $\Delta E$ between the insulating and metallic phases as a function of lattice distortion $Q$ and charge disproportionation $\Delta N $\citep{Han2018,Oleg}:
\begin{equation}
\label{eqn:energy}
\Delta E(Q,\Delta N)=\frac{kQ^2}{2}-\frac{1}{2}gQ\Delta N+E_{el}(\Delta N)
\end{equation}

The first term is the elastic energy cost of establishing the lattice distortion, the middle term is the leading symmetry allowed coupling between the structural and electronic order parameters, and the final term is the energy associated with the electronic transition. The three control parameters  are thus k, g and the combination of interaction parameters and bandwidths that determines $E_{el}(\Delta N)$. This energy formalism is general and can be applied in the context of DFT, DFT+U, DFT+DMFT and other formalisms. As the first term is meant to include all but the contribution of the correlated electrons, the value of k is independent of formalism and can be obtained by interpolation from multiple structures with varying Q from DFT alone. The lattice distortion Q leads to an on-site (Peierls) potential difference between the two inequivalent sites $\Delta_S$=gQ \citep{Subedi2015}, which is defined as the difference between the average of the on-site energies of the extended e$_g$ orbitals. This defines the second term in the energy formalism, characterized by a coupling between the electronic and lattice degrees of freedom. Finally, $E_{el}(\Delta N)$ is the energy of the correlated electrons alone and depends explicitly on the approach we use to solve the correlated problem.

To quantify the lattice distortion Q, we define the average bond disproportionation between two octahedra:
\begin{equation}
Q=\sqrt{\frac{\sum_i(l^{(i)}_{LB}-l^{(i)}_{SB})^2}{6}}   
\end{equation}
where l$^{(i)}$ are the lengths of the Ni-O bonds and LB and SB correspond to the Long Bond and Short Bond octahedra, respectively. Within our DFT+DMFT formalism, we define the electronic disproportionation $\Delta$N as:
\begin{equation}
\Delta N=N_{HF}-N_{LF}
\end{equation}
with HF = Higher Filling and LF=Lower Filling. These densities are the occupancies of the e$_g$ antibonding orbitals in our low energy model, and are simply obtained as the trace of the local density matrix on each site. When there is structural disproportionation, HF corresponds to LB and LF corresponds to SB. The occupancy of the two sites is defined within a model describing the Wannier low-energy antibonding e$_g$ bands as defined in the supplement. 

A more detailed description of the process by which we fix and determine the control parameters is given in the Appendix;  here we summarize the findings and give physical interpretations.

\begin{figure}[h]
\centering
\includegraphics[width=0.95\linewidth]{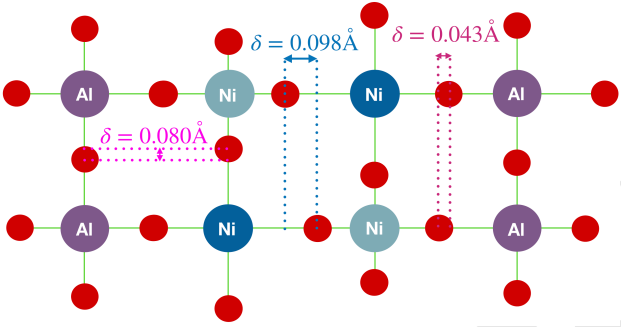}
\caption{\label{fig:bdisp}  Schematic of the bond disproportionation modes in the NNO bilayer (NNO$_2$/NAO$_2$) and its propagation into nearby NAO layers, projected on the Ni-Ni and Ni-Al direction, as discussed in the main text for a bilayer structure with the average Q=0.078 \AA - similar to the Q of the experimental bulk low T structure.}
\end{figure}

 In a previous work on bulk perovskites, the structural stiffness parameter k was found to vary only slightly as the rare earth ion was changed \citep{Oleg}. We find that heterostructuring has a stronger effect, with $k$  increasing from k=15.86eV/\AA$^2$ for bulk NNO, to 17.71eV/\AA$^2$ for the bilayer structure and 20.18eV/\AA$^2$ for the monolayer. The fundamental difference between bulk and layered systems appears at the interface between the two components of the heterostructure. A schematic of the bond disproportionation mode in the bilayer as obtained from DFT+U structural relaxations is shown in Fig. \ref{fig:bdisp}. The essential point is that the lattice distortion propagates a short distance into the counter-layer, and the stiffness to this intertwined layer-counterlayer distortion is larger than for the nickelate material alone.
 
We represent the bond disproportionation mode and its propagation in Fig \ref{fig:bdisp} for a particular Q for the bilayer. This structure is obtained through a DFT+U relaxation of a (NNO)$_2$/(NAO)$_2$ heterostructure, using a U=4eV and a c(2x2) unit cell in the xy plane, imposing ferromagnetic order on the system. This results in two pairs of inequivalent NiO$_6$ octahedra. The average bond disproportionation Q for this relaxed structure is Q=0.078\AA. This is slightly smaller than the bond disproportionation obtained from a relaxation within DFT+U with U=4eV for a bulk 20 atom unit cell for which we obtain Q=0.081\AA $ $ and smaller than the disproportionation Q=0.087\AA $ $  similarly obtained for the monolayer. Further details on the calculations and structures and the estimates of the displacements along the Ni-Ni and Ni-Al directions as pictured in the figure can be found in the supplemental information.
 
 The structural disproportionation of the bilayer octahedra is inhomogeneous: the in-plane NiO$_6$ bonds show disproportionation of about the same amplitude we would expect in the bulk. The interfacial bonds are less disproportionated as the driving force on the apical oxygen atom comes only from the Ni, the disproportionation is lower. We can then estimate the relative stiffness of the Al-O bond relative to the Ni-O bonds from the relative displacements in a simple elastic spring model, to approximately 86\% of the the stiffness of the Ni-O bonds. However, the additional energy cost per octahedron due to propagation in the NAO (or, equivalently, that NdAlO$_3$ favors a state with no bond disproportionation) leads to a higher effective stiffness per octahedron.  Finally, the out-of-plane bonds between nickelate layers in the bilayer structure disproportionate even more than the in-plane bonds, likely to compensate for the decreased interfacial disproportionation. The analysis is almost identical for the monolayer, with the exception of the nickelate inter-layer out of plane bonds which do not exist.

The structural disproportionation Q leads to an on-site potential difference between the inequivalent Ni sites. As the oxygen atoms are closer to one Ni atom than the other, this leads to a difference in electrostatic potential. Within the context of our extended e$_g$ Wannier orbitals, this can be read from the resulting Wannier Hamiltonian as the difference between the average on-site energy between the two inequivalent Ni sites:
\begin{equation}
\Delta_S=\bar{\epsilon}_{LF}-\bar{\epsilon}_{HF}  
\end{equation}
where $\bar{\epsilon}_{LF}$ is the average on-site energy of the e$_g$ orbitals on the Ni with a lower filling and $\bar{\epsilon}_{HF}$ the average for the higher filling octahedron. By analyzing multiple structures with varying amounts of structural disproportionation Q, we find that the difference in on-site potential $\Delta_S$ is linear in Q, and takes the form $\Delta_S=gQ$, with g a parameter we can determine, in agreement with previous work \citep{Oleg}. From interpolating $\Delta_S$ versus Q within DFT from multiple structures with varying Q, we can obtain a bare coupling $g^{DFT}$. As the on-site electrostatic potential difference $\Delta_S$ has to be adjusted for double counting when performing a DMFT calculation \citep{Seth2017,Oleg}(part of the on-site potential comes from Hartree interactions that appear both in DFT and DMFT), the coupling has to be adjusted within DFT+DMFT as well: $g=g^{DFT}(1+(U-\frac{5}{3}J)\chi_0)$. The value of $g^{DFT}$ is relatively constant between the bulk and heterostructured materials, and has been previously shown to be constant throughout the RNiO$_3$ family with $\chi_0=\frac{\partial \Delta N}{\partial \Delta_S}$ the electronic susceptibility as extracted from DFT. However, as $\chi_0$ is related to the inverse of the bandwidth (the occupancy changes more for the same on-site shift if the bands are narrower), the g across the materials changes slightly depending on the choice of U,J.

\subsection*{Electronic Structure}
 The dominant effect of the layering in the case of the heterostructures is electronic confinement: electron hopping is confined to be in-plane only for the NNO monolayer, and confined between the two layers for the bilayer. While the bulk orbitals have a bandwidth of 2.6eV, the \z orbital for the monolayer has a bandwidth of 1.85eV and for the bilayer 2.15eV. Two other, more minor effects appear as well. Similar to previous work \citep{BondAngles}, the bond angles from the NAO propagate into the NNO leading to straighter in-plane bond angles and  slightly higher in-plane bandwidths in the heterostructures than for bulk NNO. This leads to a \xy bandwidth of 2.72eV for the monolayer and 2.68eV for the bilayer. Previous work has shown that one can use the bond angles of the counterlayers as a control parameter to tune the metal-insulator transition temperature in nickelate heterostructures \citep{BondAngles,SROoctahedra,SroOctahedra2016}.
 
 For a lower number of layers as in this work however, the electronic confinement dominates and leads to an increased tendency to disproportionate. A third effect of heterostructuring on the electronic structure is that of the crystal-field splitting induced by the inequivalence of the bonds and the relative ionicity of the material. Finally, within the $e_g$ Wannier picture, the monolayer also shows a crystal field splitting of $\bar{\epsilon}_{x^2-y^2}-\bar{\epsilon}_{3z^2-r^2}=0.14eV$ in DFT. We've performed calculations for the monolayer with the crystal-field splitting set to 0 for U=2.1eV and found that the critical J for the spontaneous (Q=0) MIT transition line is the same as with the crystal-field splitting set to the DFT relaxed value, within an accuracy of J=0.01eV, thus showing a negligible effect (see Appendix). 
 
 The simplest way to quantify the effect of the change in bandwidth is by comparing the static electronic response to an on-site field in our $e_g$ tight-binding model, $\chi_0$ as defined previously. By reading off  $\Delta$N  versus $\Delta_S$ from multiple structures with varying amounts of structural disproportionation, we obtain: $\chi_0^{bulk}=1.16/eV$, $\chi_0^{bilayer}=1.25/eV$, $\chi_0^{monolayer}=1.39/eV$.
 
 The result of electronic confinement can be clearly seen in Figure \ref{fig:dmftq} in the curves showing $\Delta$N as a function of Q. As a response to the same structural disproportionation Q, for the same U,J parameters the monolayer is always more electronically disproportionated than the bilayer, which is always more electronically disproportionated than the bulk material ($\Delta$N$_{monolayer}$>$\Delta$N$_{bilayer}$>$\Delta$N$_{bulk}$). Further, there is a range of U, J parameters (which the middle plot in the figure samples) for which the heterostructures can be insulating, even in the absence of any structural disproportionation (Q=0)).

\begin{figure}[h]
\includegraphics[width=1.1\linewidth]{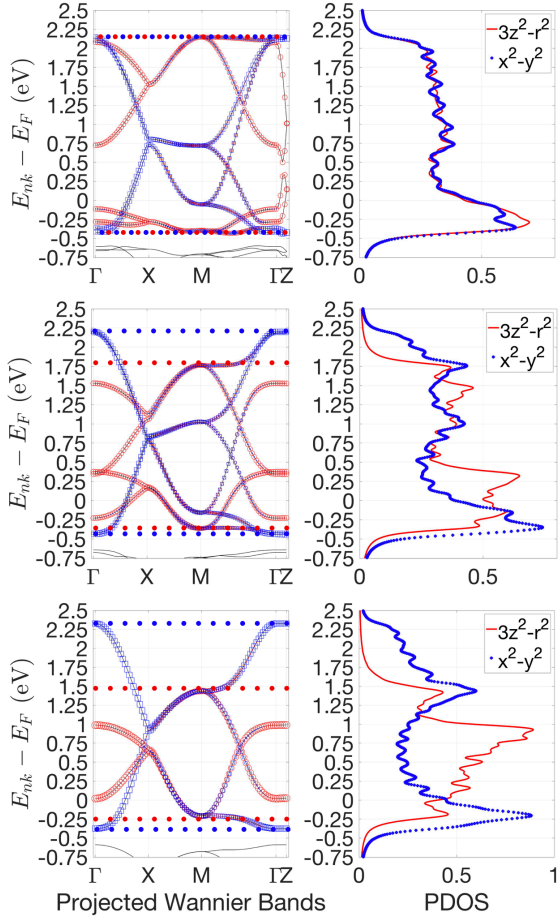}
\caption{\label{fig:HetStruct} Projected density of states of low-energy e$_g$ Wannier bands for GGA-relaxed structures for bulk (top), bilayer (middle) and monolayer (bottom) structures. Dotted horizontal lines show approximate cutoff for determining bandwidths mentioned in main text. Confinement greatly reduces the bandwidth of the \z orbital, however the bond angle propagation leads to a slightly wider bandwidth of the \xy orbital.}
\end{figure}

We have then found three main effects of heterostructuring on the electronic structure. The effect of layer confinement strongly lowers the kinetic energy of the electrons and favors an insulating state, with its effect primarily on the \z orbital. The bond angle propagation leads to a small effect in the opposite direction, primarily on the \xy band. Finally, the crystal field splitting is only significant in the monolayer, however it does not affect the electronic transition.


\subsection*{Equilibrium bond and electronic disproportionation from total energy model}
We now turn to determining equlibrium points in the energy functional from equation \ref{eqn:energy}. Stationarity of $\Delta E$ with respect to variations in $\Delta N$ and Q implies the two equilibrium conditions:
\begin{equation}
\label{eqn:Q}
0=kQ-\frac{1}{2}g\Delta N
\end{equation}
and:
\begin{equation}
\label{eqn:deltaN}
0=-\frac{1}{2}gQ+\frac{\partial E_{el}(\Delta N)}{\partial \Delta N}
\end{equation}
Equation \ref{eqn:Q} gives Q as a function of $\Delta N$ 
 However, its meaning is very simple: for a particular value of the electronic disproportionation $\Delta N$ one can obtain the equilibirum structural displacement Q of the oxygen atoms as a result of the resulting electrostatic forces. Equation \ref{eqn:deltaN} gives $\Delta N$ as a function of Q, as obtained via the density functional plus dynamical mean field theory (DFT+DMFT) method.

Combining the two, we have an equation of state \citep{Oleg}:
\begin{equation}
\label{eqn:state}
\frac{2k}{g}Q=\Delta N[Q]
\end{equation}

In practical terms we can use this equation in a very simple manner: using the stiffness k and coupling g obtained from the interpolation from DFT calculations and adjusting g for double counting, we can obtain the equilibrium Q for a particular $\Delta$ N as Q=g$\Delta N$/2k. Separately, we obtain the equilibrium $\Delta N$ as a function of Q from explicity DFT+DMFT calculations rather than from equation \ref{eqn:state}. The effect of Q is simulated by applying on-site terms to the Q=0 Hamiltonian, namely $\Delta_S/2$ to simulate the short bond octahedron and $-\Delta_S/2$ the long bond octahedron, where $\Delta_S$ is obtained from Q simply by multiplying $\Delta_S=g Q$. Single-shot DMFT calculations are then performed on the resulting Hamiltonian to obtain $\Delta N$. The intersection of the functions $\Delta $N[Q] and Q[$\Delta$N] then determine equilibrium solutions for the material.

\begin{figure}[h]
\includegraphics[width=1.0\linewidth]{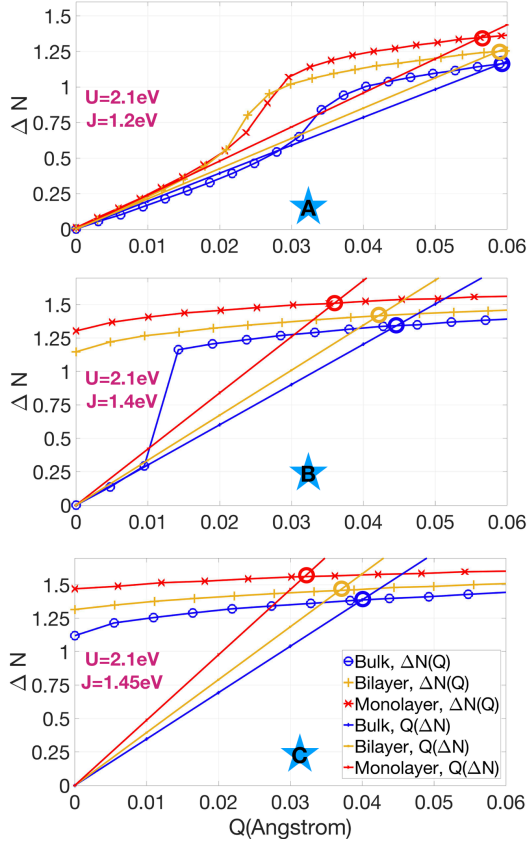}
\caption{\label{fig:dmftq} $\Delta N$ versus bond disproportionation Q within DFT+DMFT for bulk, bilayer and monolayer structures as well as Q versus $\Delta N$ lines from the total energy model calculation in different areas of the phase space as determined by equation \ref{eqn:Q} and equation \ref{eqn:deltaN} via DFT+DMFT as described in the main text. The thick circles mark the intersections that respect the equation  of state \ref{eqn:state}. The panels marked with A,B,C correspond to the points marked with stars in the phase diagram in Fig. S1 in the supplemental information.}
\end{figure}

The $\Delta N(Q)$ relation is shown in Figure \ref{fig:dmftq} as large symbols connected by lines for $U=2.1eV$ and three $J$ values. For the smallest $J$  value neither the bulk nor the superlattice materials show a spontaneous disproportionation at $Q=0$; for small $Q$ there is a regime in which the disproportionation is linear in $Q$ and the solution remains metallic. Above a particular $Q$ there is a very rapid crossover to an insulating solution with a $\Delta N$ which is large and only weakly dependent on $Q$. In the insulating regime the monolayer has a  larger disproportionation than the bulk, with the bilayer in between.  For an intermediate $J$, the monolayer and bilayer exhibit a spontaneous disproportionation at $Q=0$ but the bulk material exhibits a $Q$-driven first order transition. At the larger $J$ all three systems spontaneously distort at $Q=0$. 

Also shown in Figure \ref{fig:dmftq} are straight lines corresponding to the  Q($\Delta N$) relation from equation \ref{eqn:Q}.  The intersection of these lines with the DMFT $\Delta N(Q)$ curves defines the actual values of $\Delta N$ and Q. We see from the relative positions of the intersections that $Q_{monolayer}<Q_{bilayer}<Q_{bulk}$ and $\Delta N_{monolayer}> \Delta N_{bilayer} > \Delta N_{bulk}$. From an electronic point of view the monolayer and bilayer are more disproportionated ($\Delta N$ is larger) as $\Delta N$ does not depend strongly on Q,  however the higher stiffness of the heterostructures leads to a lower Q. Further, as shown in the middle figure in \ref{fig:bdisp}, there is a range of U,J for which the heterostructures will stay insulating even at a very small Q while the bulk becomes metallic. 

The relative roles of the lattice and electronic structure are easily disentangled from the above. First, the electronic disproportionation has a first order transition, followed by a very slowly varying $\Delta N$ in the insulating phase. Assuming $\Delta N$ is nearly constant in the insulating phase $\Delta N\approx N_{insulating}$, Q is then set by optimizing the structure Q[$\Delta N$] in equation [4] as approximately $Q\approx \frac{g \Delta N_{insulating}}{2k}$. This allows the seemingly paradoxical solutions with the amplitude of Q and $\Delta N$ showing opposite trends between the bulk and heterostructure. If we assume that the experimentally obtained metal-insulator transition temperature is more strongly correlated to $\Delta N$ than Q, while the XAS spectra splitting is more strongly correlated with Q, we can thus explain the seemingly paradoxical results in previous work \citep{Disa2017}.

One of the signatures associated with the bond disproportionated phase of the RNO nickelates is an increased peak-prepeak splitting of the XAS Ni L$_3$ edge, which in the monolayer and bilayer were found to be in between the values of the bulk disproportionated and undisproportionated structures throughout the insulating temperature range scanned. Consistent with this result, we find that the predicted value of the structural disproportionation of the monolayer is lower than the bulk, with the bilayer in between the two. Further, XAS integration of the monolayer in-plane and out-of-plane Ni L$_3$ edge has found an orbital polarization of ~8\% favoring the \z orbital. Within our insulating solutions we find that orbital polarization is strongly suppressed ( <2\%) however we consistently find that the long bond site has an orbital polarization of 5-8\% in a direction consistent with experiment, while the LB site is orbitally polarized of about the same magnitude but in the opposite direction. This suggests that the XAS spectra may sample primarily the LB, however further theoretical and experimental work is needed.

\subsection*{Conclusion and Outlook}
Using a combination of DFT+DMFT and many-body theory we have elucidated the relative importance of lattice and electronic effects in heterostructured materials. We have found that the higher lattice energy cost in the heterostructured materials decreases the structural signatures of the symmetry-broken phase within the correlated material going through an MIT but that the distortion associated with it can propagate into the epitaxial layer. We have found that, as the effect of interactions is increased in a layer-confined structure, electronic disproportionation can be higher despite lower structural distortions in a heterostructure. Through comparison with experiment \citep{Disa2017}, our study suggests that the electronic disproportionation is more likely to be correlated to the metal insulator transition temperature than the structural disproportionation, which is suppressed by the higher structural stiffness of the material. At the same time, our work suggests that the structural disproportionation is more strongly connected to the XAS splitting observed experimentally, likely via the induced on-site electrostatic potential difference.

These general results can be used both to understand other similar heterostructures (for example LaNiO$_3$/LaAlO$_3$) as well as to design new materials. Our analysis of the bond disproportionation mode on the interfacial structure in this class of materials as well as in related classes of materials (vanadates, manganites etc) can be studied both theoretically and experimentally. The combination of bond angles, confinement and relative structural stiffness can be used to fine tune metal-insulator transition temperatures. Based on the methodology in this work and previous work \citep{Oleg,Han2018}, future work involving DFT+DMFT, DFT+U studies and model calculations, can address the relative roles of lattice and electronic disproportionation.

\subsection*{Methods}
For our calculations, we use structures obtained from fully relaxed DFT+U calculations \citep{Anisimov1997} and impose 0\% strain relative to the theoretical DFT bulk NNO lattice constant on the heterostructures. We use Quantum Espresso, ultrasoft pseudopotentials, either from the GBRV or generated using the Vanderbilt ultrasoft pseudopotential generator as described in previous work \citep{Anisimov1997, QE,ultrasoft, KevinPSP, Alex2014}  and benchmark our results against experimental bulk structures. The disproportionated structures have two inequivalent Ni sites, one with relatively long Ni-O bonds ('LB') and one with relatively short Ni-O bonds ('SB'). We define the structural order Q as:
\begin{equation}
Q=\sqrt{\frac{\sum_i(l^{(i)}_{LB}-l^{(i)}_{SB})^2}{6}}   
\end{equation}
where l$^{(i)}$ are the lengths of the Ni-O bonds. 

For each structure we then perform a self consistent DFT calculation and fit the bands arising from the frontier \eg  orbitals using maximally localized Wannier functions as implemented in Wannier90 \cite{wannier90,Marzari2012}. Bands for representative structures near the Fermi level and their Wannier fits are shown in Figure \ref{fig:HetStruct}.
The parameter g in equation \ref{eqn:Q} is defined in terms of the on-site energy difference $\Delta_S=gQ$ entering our DMFT calculations.
In our one-shot DMFT, g is corrected from the DFT value by a double counting term \citep{Oleg,Seth2017}, so $g=g^{DFT}(1+(U-\frac{5}{3}J)\chi_0)$. 

The Wannier fits define a low energy tight binding model to which we add standard Slater-Kanamori interactions and solve using dynamical mean field theory (using the triqs library\citep{TRIQS}, ct-hyb solver\cite{CTHYB} and dfttools \citep{DFTTOOLS} interface)  with the two inequivalent Ni treated as different embedded atoms.

The parameter k is the stiffness to lattice distortions at fixed $\Delta N$. We argue, following \citep{Oleg,Han2018} that since the stiffness comes from the full electronic structure at fixed $\Delta N$, the frontier orbitals play a relatively minor role and for the purpose of calculating k may be treated at the DFT level.
We therefore obtain k from the dependence of the DFT energy on Q $\frac{\partial E_{DFT}}{\partial Q}=cQ$. However in the DFT calculations $\Delta N$ is relaxed at each Q . Referring to equation \ref{eqn:Q} we have on the DFT level (and noting the stationarity with respect to $\Delta N$):
\begin{equation}
\label{eqn:c}
    cQ=\frac{\partial E_{DFT}}{\partial Q}=kQ-\frac{1}{2}g^{DFT}\Delta N^{DFT}(Q)
\end{equation}
In the linear response regime, which accurately describes the DFT results for all structures we considered, we find:
\begin{equation}
    \frac{cQ^2}{2}=E_{DFT}(Q)=(\frac{k}{2}-\frac{1}{4}(g^{DFT})^2\chi_0)Q^2
\end{equation}
We can then extract c from the energy of continuously varying structures with different Q. Combining this with knowledge of $g^{DFT}$ and $\chi_0$ as described below we can then obtain the stiffness k.
The parameter $g^{DFT}$ is defined in terms of the average on site energy $\Delta_S^{DFT}$ obtained from our Wannier fits to DFT band structures as $g^{DFT}=\frac{\Delta_S^{DFT}}{Q}$ and $\Delta N(Q)$ is obtained from the occupancy difference of the Wannier orbitals and is found to be linear in Q, $\Delta N=\chi_0 g^{DFT}Q$. This relation defines the on-site susceptibility $\chi_0$. $g^{DFT}$ can be read off from the on-site energy difference and is nearly identical for all three materials, namely 2.89ev/\AA $ $  for bulk and 2.972ev/\AA $ $ for the bilayer and 2.962ev/\AA $ $ for the monolayer. This means that a similar movement of the ions leads to a similar change of electrostatic potential, which is something we would expect as the local environment is similar.

\acknow{We thank  Sohrab Ismail-Beigi, Jean-Marc Triscone, Hugo U.R. Strand, Manuel Zingl, Alexander Hampel and Claude Ederer for helpful conversations and Nick Carriero and the Scientific Computing Core division of the Flatiron Institute for invaluable technical assistance. The Flatiron Institute is a division of the Simons Foundation. AG acknowledges the support of the European Research Council (ERC-319286-QMAC). OEP acknowledges funding from Forschungsf\"orderungsgesellschaft (FFG) COMET program IC-MPPE (project No 859480)}

\showacknow{} 

\subsection*{References}
\bibliography{Mendeley}

\end{document}